\documentclass[final,5p,times,twocolumn]{elsarticle}

\usepackage{graphicx}
\usepackage{multirow}
\usepackage{wrapfig}
\usepackage{array}

\usepackage{tabularray}
\usepackage{tabularx}
\usepackage{rotating}
\usepackage{makecell}
\usepackage{caption}
\usepackage{subcaption}
\usepackage{booktabs}
\usepackage{csquotes}

\usepackage[utf8]{inputenc}
\usepackage{amsmath}
\usepackage{amsthm}
\usepackage{amsfonts}
\usepackage{epsfig}
\usepackage{psfrag}
\usepackage{pstricks}
\usepackage{algorithm}
\usepackage{stfloats}
\usepackage{cuted}

\usepackage{algorithmicx}
\usepackage{algpseudocode}  
\floatname{algorithm}{Algorithm}

\graphicspath{{./Figures/}}

\usepackage{amssymb}
\usepackage{bm}
\usepackage{hyperref}

\usepackage{booktabs}
\usepackage{array}
\usepackage{ragged2e}
\newcolumntype{P}[1]{>{\RaggedRight\arraybackslash}p{#1}}

\hypersetup{
            colorlinks=true,
            linkcolor=blue,
            anchorcolor=blue,
            citecolor=blue
            }

\biboptions{comma,sort&compress}

\usepackage{enumitem}
\setlist{nosep,topsep=-\parskip}

\usepackage{ulem}

\journal{CAD}

\begin{document}

\begin{frontmatter}

\title{{Bringing Attention to CAD: Boundary Representation Learning via Transformer}}

\author[]{Qiang Zou\corref{cor}}\ead{qiangzou@cad.zju.edu.cn}
\author[]{Lizhen Zhu}

\cortext[cor]{Corresponding author.}
\address{State Key Laboratory of CAD$\&$CG, Zhejiang University, Hangzhou, 310058, China}

\begin{abstract}
The recent rise of generative artificial intelligence (AI), powered by Transformer networks, has achieved remarkable success in natural language processing, computer vision, and graphics. However, the application of Transformers in computer-aided design (CAD), particularly for processing boundary representation (B-rep) models, remains largely unexplored. To bridge this gap, we propose a novel approach for adapting Transformers to B-rep learning, called the Boundary Representation Transformer (BRT). B-rep models pose unique challenges due to their irregular topology and continuous geometric definitions, which are fundamentally different from the structured and discrete data Transformers are designed for. To address this, BRT proposes a continuous geometric embedding method that encodes B-rep surfaces (trimmed and untrimmed) into Bézier triangles, preserving their shape and continuity without discretization. Additionally, BRT employs a topology-aware embedding method that organizes these geometric embeddings into a sequence of discrete tokens suitable for Transformers, capturing both geometric and topological characteristics within B-rep models. This enables the Transformer’s attention mechanism to effectively learn shape patterns and contextual semantics of boundary elements in a B-rep model. Extensive experiments demonstrate that BRT achieves state-of-the-art performance in part classification and feature recognition tasks.
\end{abstract}


\end{frontmatter}


\section{Introduction}
\label{sec:introduction}
Extracting semantic information from computer-aided design (CAD) models via deep learning is a critical step toward developing next-generation intelligent CAD systems~\cite{zou2024intelligent}. CAD models are typically represented by the boundary representation (B-rep) scheme, a hierarchical collection of boundary elements of vertices, edges, loops, faces, and shells~\cite{zou2023variational}. Their irregular topology (i.e., connectivity among boundary elements) and continuous geometric definitions (i.e., parametric curves and surfaces) pose unique challenges distinct from the structured formats for which most deep neural networks are designed, such as sequences (1D text) or grids (2D images).

A common approach to addressing these challenges involves converting B-rep models into intermediate representations, such as point clouds, voxels, or multi-view images, to allow direct application of existing deep networks~\cite{qi2017pointnet++,zhang2022machining,zhang2018featurenet,peddireddy2021Identifying,su2015multi,shi2020novel}. However, these methods often compromise topological information, which is critical for preserving the integrity of B-rep models and supporting downstream applications. Recent efforts have turned to graph neural networks (GNNs) to explicitly process topological information~\cite{jayaraman2021uv,colligan2022hierarchical,lee2023brepgat,zhang2024brepmfr,wu2024aagnet}. Despite the improved performance, these methods still rely on converting continuous geometric data into discrete forms. No universal discretization parameters suit all cases and consequently, tedious manual tuning is required, limiting these methods' applicability.

Recently, Transformer networks have emerged as the dominant architecture in natural language processing, computer vision, and computer graphics~\cite{vaswani2017attention,dosovitskiy2020image,guo2021pct}. There are good reasons to extend Transformers to the CAD domain, particularly for B-rep models. First, Transformers serve as the backbone of the recent generative AI technologies, such as ChatGPT~\cite{brown2020language}. Adapting Transformers for B-rep learning allows the CAD domain to stay aligned with this AI trend. Second, the attention mechanism of Transformer networks excels at capturing global, complex relationships within input data, and this makes them particularly well-suited for modeling the intricate geometric and topological patterns among non-adjacent yet functionally related surfaces within B-rep models~\cite{li2020survey}, which are challenging to handle with traditional neural networks like GNNs.

This paper presents an approach, called Boundary Representation Transformer (BRT), for learning B-rep models using Transformers. Unlike languages or images, which exhibit regular structures, B-rep models are characterized by their irregular topology~\cite{zou2019push}. Moreover, while languages and images are inherently discrete in their content, B-rep models are continuous in their geometry (e.g., B-spline curves and surfaces~\cite{zou2025splinegen}). Consequently, extending Transformers originally designed for sequential and discrete data to B-rep models is no trivial matter. The fundamental challenge here is transforming irregular, continuous B-rep models into a sequence of structured tokens that Transformers can effectively process. (A token is a small, meaningful unit of raw data that AI models can understand, for example, a word~\cite{dosovitskiy2020image}.)

To effectively tokenize B-rep models, BRT introduces a novel geometric embedding method that maps each boundary element into a high-dimensional vector space, where the vector captures the geometric characteristics of the boundary element. Unlike existing approaches, this method operates completely in the continuous domain, eliminating the need for any discretization. This is achieved by decomposing a surface (trimmed or untrimmed) into a collection of Bézier triangles while preserving the surface shape. Bézier triangles are advantageous because they are continuous, offer greater flexibility in shape representation, and most importantly, have a regular structure---always having a fixed number of control points for a given degree~\cite{farin1986triangular}.

After obtaining the geometric embeddings, BRT further applies a topology-aware embedding method to organize all the geometric embeddings into face-based tokens for input into Transformers. It leverages the hierarchical relationships (e.g., vertex-edge relationships), cyclic relationships (e.g., edge-loop relationships), and parallel relationships (e.g., face-face adjacency relationships) among the boundary elements to guide the token organization. (It is not straightforward for GNNs to handle these topological relationships simultaneously.) As such, the resulting tokens capture not only the geometric characteristics but also the topological patterns of the boundary elements within B-rep models. This enables more comprehensive and context-aware tokenization of B-rep models compared to existing methods. As a result, it can provide a solid foundation for Transformers to learn meaningful representations that capture the semantics within B-rep models.

The main contributions of this paper are as follows:
\begin{enumerate}
    \item A novel tokenization method for B-rep models to be directly processed by the (sequential) Transformer architecture. In contrast, prior works rely on attention-based GNNs, which are limited in capturing global relationships due to their localized message-passing mechanisms. Our tokenization approach also aligns with the broader trend of 'tokenizing everything" in NLP and CV, opening the door for integration with these fields. To the best of our knowledge, this is the first approach to integrate B-rep models and (sequential) Transformers.
    \item A new geometric embedding method that can work entirely in the continuous domain, which leads to the advantage of capturing finer geometric details. Previous methods are consistently based on the discretization of B-rep surface patches into points, meshes, or voxels, which inevitably introduce errors.
    \item A topology-aware embedding method for more comprehensive and context-sensitive tokenization of B-rep models compared to existing approaches.
\end{enumerate}

The remainder of this paper is organized as follows. Sec.~\ref{sec:related_work} reviews the literature, Secs.~\ref{sec:methods} elaborate on the components of BRT. Validation of the method using a series of examples and comparisons can be found in Sec.~\ref{sec:results}, followed by conclusions in Sec.~\ref{sec:conclusion}.

\section{Related Work}
\label{sec:related_work}
With rapid advancements in deep learning, 3D geometric modeling is shifting from traditional rule-based methods to data-driven approaches. B-rep modeling follows this trend, with development broadly falling into two categories: conversion-based and direct methods. The former converts B-rep models into intermediate representations, such as point clouds, voxels, or multi-view images, allowing the direct use of existing deep learning networks from computer vision and graphics. The latter focuses on directly modeling the topological aspects of B-rep models using neural networks like GNNs. The following summarizes these two lines of research, as well as the application of Transformer networks to 3D modeling, which is closely related to this work.

\textbf{Conversion-Based Methods.} Effective learning algorithms have been developed for discrete 3D models such as point clouds, meshes, and voxels. Notable examples include PointNet/PointNet++~\cite{qi2017pointnet,qi2017pointnet++} and Dynamic Graph CNN~\cite{wang2019dynamic} for point clouds, MeshCNN~\cite{hanocka2019meshcnn}, MeshNet~\cite{feng2019meshnet}, and MeshWalker~\cite{lahav2020meshwalker} for meshes, and O-CNN~\cite{wang2017cnn} for voxels, among others. Recent works in these areas (e.g.,~\cite{zhao2021pointtransformer,yu2022point,li2022meshformer,hu2022subdivision,mao2021votr}) have further advanced the state of the art. In this regard, a natural choice for handling B-rep models is to convert them into these discrete representations, as demonstrated by methods like FeatureNet~\cite{zhang2018featurenet}, Mesh-Faster RCNN~\cite{jia2023machining}, and ASIN~\cite{zhang2022machining}. In addition to 3D discrete representations, some approaches convert 3D objects to 2D views, allowing traditional 2D CNNs to be applied directly~\cite{shi2015deeppano,liu2016ssd,shi2020novel}. While these conversion-based methods have been effective in certain applications, they compromise topological information (i.e., connectivity among boundary elements), which is crucial for maintaining the integrity of B-rep models and supporting downstream tasks, limiting their overall applicability.

\textbf{Direct Methods.} These methods aim to directly learn from B-rep models without relying on intermediate representations. While presented in various forms, common to them is the use of GNNs to process the model's topology~\cite{cao2020graph,jayaraman2021uv,lambourne2021brepnet,colligan2022hierarchical,lee2023brepgat,zhang2024brepmfr,wu2024aagnet}. This concept was first introduced by Cao et al.~\cite{cao2020graph}, although their work was restricted to planar surfaces. Subsequent advancements expanded this approach to quadratic and freeform surfaces, as seen in UV-Net~\cite{jayaraman2021uv} and Hierarchical CADNet~\cite{colligan2022hierarchical}. These methods discretize surfaces into point clouds or meshes and use existing learning algorithms (e.g., CNNs) to encode them as GNN nodes. In addition to discretization, heuristic descriptors (such as face type, face genus, and face-face concavity/convexity) have also been employed to encode faces for use as GNN nodes, e.g., BRepGAT~\cite{lee2023brepgat}, BRepMFR~\cite{zhang2024brepmfr}, and AAGNet~\cite{wu2024aagnet}. While direct methods outperform conversion-based approaches, they often depend on surface discretization or simplistic heuristic descriptors, which require manual tuning to achieve satisfactory results and limit their applicability.

\textbf{Transformer in 3D Learning.} Transformer networks, first introduced by Vaswani et al.~\cite{vaswani2017attention}, utilize attention mechanisms to allow each token to focus on the most relevant parts of the input, effectively capturing long-range dependencies and contextual relationships. Inspired by their success in NLP, researchers have explored their potential for processing 3D models. For point clouds, methods such as Point Transformer~\cite{zhao2021pointtransformer}, Point Cloud Transformer (PCT)~\cite{guo2021pct}, and Residual Attention Network~\cite{ma2022rethinking} exploit the self-attention mechanism to capture local and global geometric features. In mesh processing tasks like segmentation and reconstruction, MeshFormer~\cite{li2022meshformer} and Mesh Graphormer~\cite{lin2021mesh} leverage attention to model complex topological relationships. For voxel-based representations, Transformers have been applied to both dense and sparse voxels. Examples include Voxel Transformer~\cite{mao2021votr} for learning on dense grids and OctFormer~\cite{wang2023octformer} for learning octree voxels. The application of Transformers in these domains has led to notable performance improvements over traditional architectures such as CNNs and GNNs.

The developments outlined above highlight the effectiveness of Transformers in processing 3D models. However, their application to B-rep models remains unexplored due to the complex topological structures and continuous geometric data within B-rep models. The proposed BRT attempts to bridge this gap by addressing these challenges through novel continuous geometric embedding and topology-aware embedding methods.

\section{Methods}
\label{sec:methods}
BRT consists of two key components: geometric embedding and topological embedding. Geometric embedding encodes vertices, edges, and faces into high-dimensional latent vectors that capture their geometric properties. Topological embedding then aggregates these embeddings based on B-rep topology, forming tokens for input into Transformers. Since topological embedding defines the overall structure while geometric embedding handles finer details, we first present the topological embedding and then the geometric embedding.

\begin{figure*}[t]
    \centering
    \includegraphics[width=\linewidth]{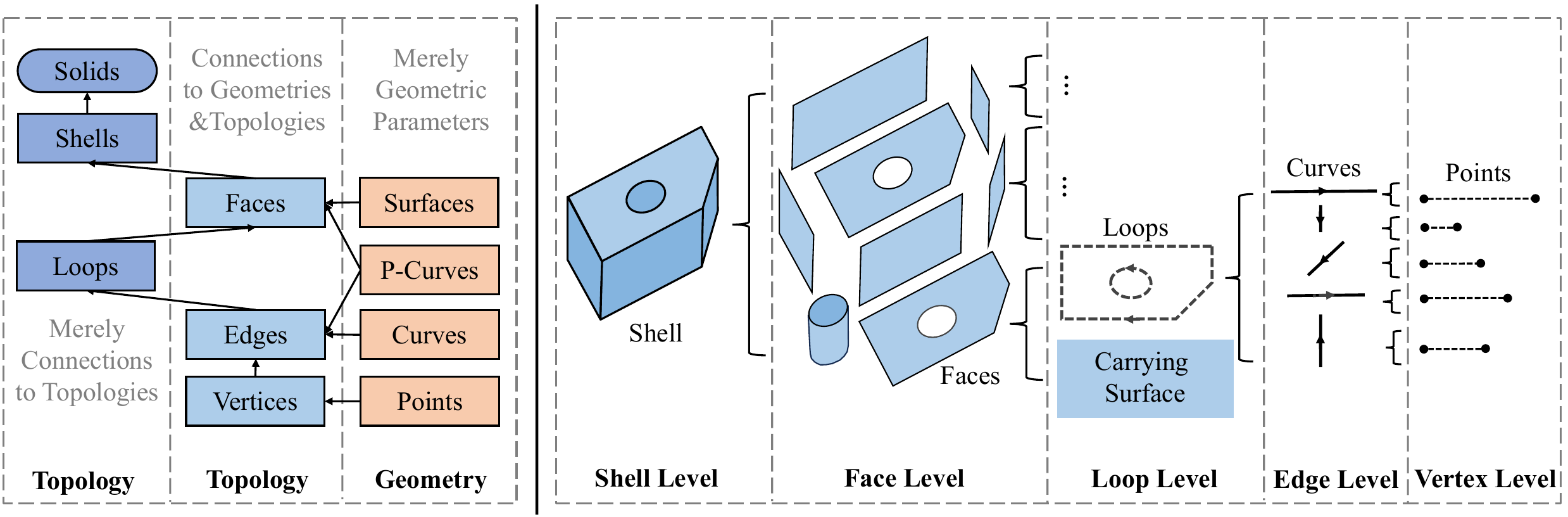}
    \caption{The relationship of geometric and topological elements in a B-rep model (left) and their illustration (right).}
    \label{fig:b-rep}
\end{figure*}

\subsection{Topological Embedding of B-rep Models}
\label{sec:topological-embedding}
Topological embedding organizes the geometric embeddings of vertices, edges, and faces based on their topological connectivity (e.g., linking edges into loops) to generate face-based tokens. Using the B-rep topology to guide this organization, we can ensure that each face token encapsulates not only its own geometric characteristics but also those of all boundary elements topologically related to it. 

The concept of B-rep originates from solid modeling, where a solid is mathematically defined as a bounded, closed, regular, and semi-analytic subset of $\mathbb{R}^3$, commonly known as an r-set~\cite{zou2023variational}. While this definition includes most engineering products, it also permits non-manifold models (which are non-manufacturable). Therefore, solids are preferred to be r-sets with 2-manifold boundaries. For formal definitions of regularity, semi-analyticity, and manifold, refer to~\cite{zou2023variational} and the references therein.

A solid model is a computer representation of a solid, and a B-rep model is a type of solid model that defines a solid by specifying the boundary between the solid and the surrounding void. As shown in Fig.~\ref{fig:b-rep}, it comprises topological elements (vertices, edges, loops, faces, and shells) and their connections, along with geometric definitions of these elements (i.e., points, curves, and surfaces). A vertex corresponds to a point; an edge is a curve segment bounded by two end vertices; a loop forms a closed circuit of edges; a face represents a bounded surface enclosed by loops; and a shell is a collection of connected faces.

\begin{figure*}[t]
    \centering
    \includegraphics[width=\linewidth]{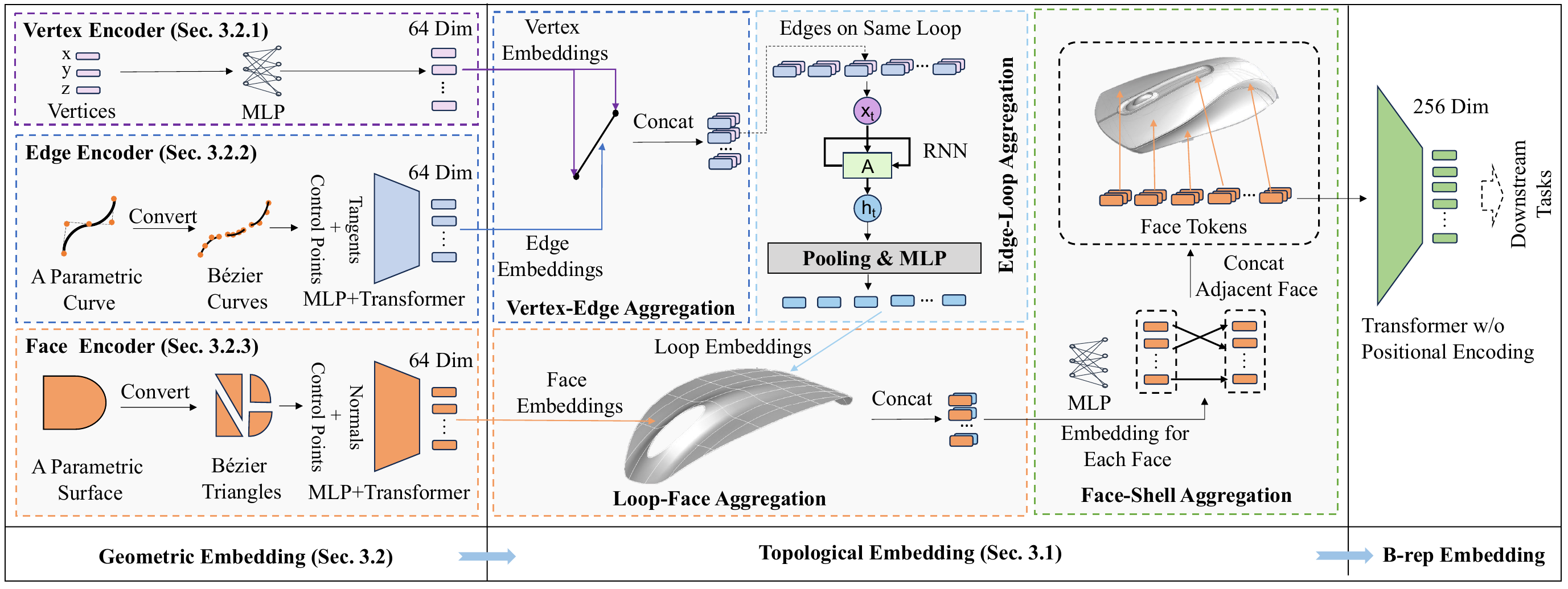}    
    \caption{Overall framework of the BRT model.}
\label{fig:topology-embedding}
\end{figure*}

A B-rep model encompasses various topological relationships, including hierarchical (e.g., vertex-edge connections), cyclic (e.g., edges forming a loop), and parallel (e.g., adjacent faces within a shell). In this work, we construct a multi-level directed graph that accurately represents the topology of the solid model. Specifically, the top level consists of faces; each face is attached to its boundary loops. Each loop, again, is composed of edges, which are all inherently directed with the start vertex and the end vertex attached to it. Different topological embedding methods are provided for these relationships, as detailed below and illustrated in Fig.~\ref{fig:topology-embedding}. In the following discussion, we assume that the geometric embeddings of vertices, edges, and faces have already been made available, allowing us to focus exclusively on the topological aspects of B-rep embedding.

\textbf{Vertex-Edge Aggregation.} 
The connections between vertices and edges follow a regular pattern: each edge connects to two vertices\footnote{A special case arises when an edge forms a closed loop, such as a circle with a single endpoint. In this case, the endpoint is duplicated to maintain the one-edge-two-vertex pattern.}. The two vertices are designated as the start vertex and the end vertex of the edge, which has a direction that follows the loop to which it belongs. Given this structured connectivity, we aggregate information by concatenating the embeddings of the two vertices with the edge embedding. Specifically, for an edge $ e_k $ connected to the start vertice $ v_i $ and the end vertice $ v_j $, with embeddings $ E(e_k) $, $ E(v_i) $, and $ E(v_j) $, respectively, the aggregation is done as follows:
\begin{equation}
     E(e_k) \longleftarrow \text{Concat}(E(e_k), E(v_i), E(v_j)).
\end{equation}

A practical note should be made here. Some might argue that combining vertex embeddings with edge embeddings is unnecessary since the underlying curve segment already encodes endpoint information. However, explicitly incorporating vertex embeddings allows edges sharing the same vertex to exchange information more effectively. Without this, the network would need to learn these connections on its own, adding unnecessary complexity to the learning process.

\begin{figure}[htbp]
    \centering
    \includegraphics[width=\linewidth]{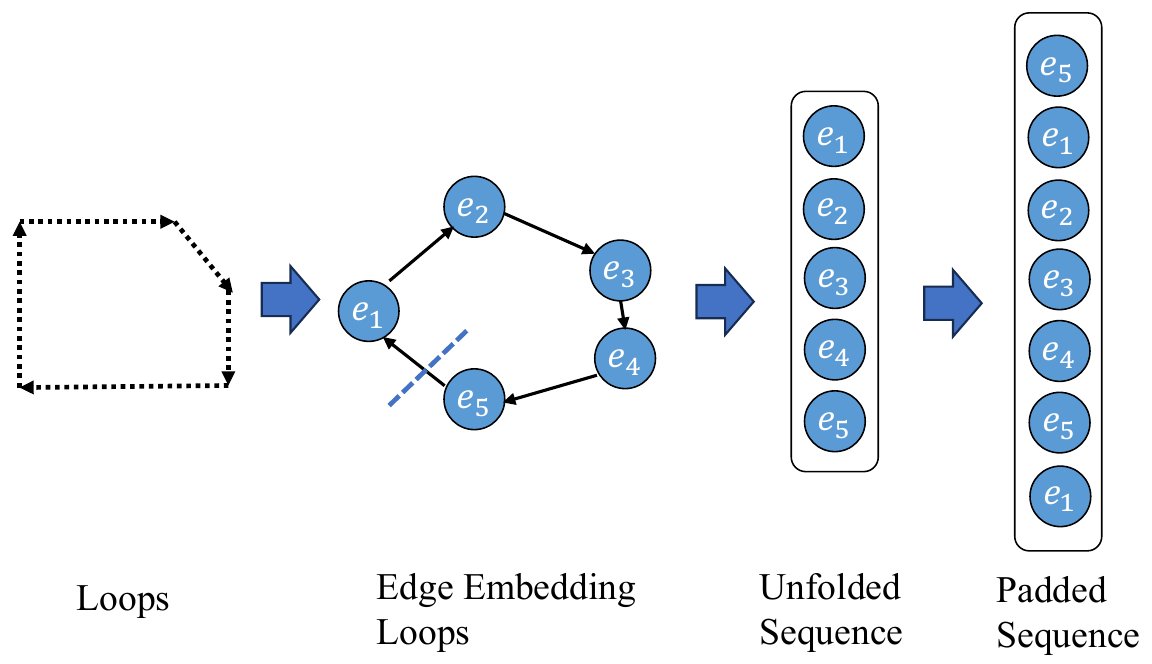}
    \caption{The process of unfolding and padding a loop.}
    \label{fig:edge-loop-aggregation}
\end{figure}

\textbf{Edge-Loop Aggregation.} 
Unlike the previous aggregation, the number of edges in a loop varies. Also, these edges form a closed circuit, meaning there is no inherent ordering among them, although each edge has an immediate previous and next edge. To address this, we use a recurrent neural network (RNN) to aggregate the edge embeddings. However, loops do not have a distinct start or end, which is typically required to run RNNs. To resolve this, we break the loop at a random edge and unfold it into a linear format that the RNN can process; padding is then added before the start and after the end (Fig.~\ref{fig:edge-loop-aggregation}). This padding process allows the RNN to focus on the loop's cyclic connections without being biased by a particular breaking point. The RNN then takes this padded sequence as input.

Mathematically, let the unfolded edge sequence be $ \{e_1, e_2, \dots, e_n, e_1\} $. The padded version is then: 
\begin{equation}
    \{e_n, e_1, e_2, \dots, e_n, e_1, e_2\}.
\end{equation}

Applying an RNN to this padded sequence is represented by the following equation:
\begin{equation}
    h_i = \tanh((W_{he} E(e_{i}) + W_{hh} h_{i-1}) W_h)
\end{equation}
where $ h_i $ is the $i$-th hidden state of the RNN, $ W_{he} $ and $ W_{hh} $ are the weights used to convert the edge embeddings and the previous hidden state into a new hidden state representation, respectively. The matrix $ W_h $ is used to transform the combined representation into the final hidden state.

\textbf{Loop-Face Aggregation.} 
Loops define the boundaries of faces, with each face having one outer boundary loop and a variable number of inner boundary loops (or holes). Since there is no inherent order among the inner boundary loops, we first apply mean and max pooling to their embeddings to generate two aggregated embeddings that capture their overall characteristics. We then use an additional multilayer perceptron (MLP) to combine these two aggregated embeddings with the embedding of the outer boundary loop, producing a unified loop embedding, which is finally concatenated with the face embedding. Specifically, the MLP consists of two layers, featuring a hidden state of size 256 and producing an embedding of size 64.

Let a face be denoted by $f_i$, its outer boundary loop be $l_0$, and inner boundary loops be $\{l_1, \dots, l_n\}$. The above aggregation process can be mathematically expressed as:
\begin{align}
    E(f_i) \longleftarrow &\text{Concat}(E(f_i), \nonumber\\
    &MLP(Concat(E(l_0), \nonumber\\
    &\text{MeanPooling}(E(l_1), \dots, E(l_n)), \nonumber\\
    &\text{MaxPooling}(E(l_1), \dots, E(l_n)))))
\end{align}
where $E(\cdot)$ denotes the embedding of a loop or a face.

\textbf{Face-Shell Aggregation.} 
A shell is formed by a group of connected faces. In most cases, a B-rep solid model simply consists of a single closed shell. However, shells can vary in many ways, such as the number of faces, disjoint components, or internal voids. Since our goal is to achieve a face-based tokenization of B-rep models, we aggregate each face's embedding with those of its 1-ring neighboring faces. This aggregation captures local contextual information while ensuring overlapping embeddings among faces. Such overlap is particularly beneficial for the attention mechanism in Transformers to effectively model global contextual information across the entire shell.

Specifically, given a face $f_i$ and its neighboring faces $\{f_1, \dots, f_n\}$, we apply mean and max pooling to their embeddings to generate two aggregated embeddings that capture their overall characteristics, following a similar approach to loop-face aggregation. We then use a MLP (two layers of sizes 512 and 64) to combine these aggregated embeddings into a unified neighboring face embedding. Finally, we concatenate this with the embedding of the central face $f_i$:
\begin{align}
    E(f_i) \longleftarrow &\text{Concat}(E(f_i), \nonumber\\
    &MLP(Concat( \nonumber\\
    &\text{MeanPooling}(E(f_1), \dots, E(f_n)), \nonumber\\
    &\text{MaxPooling}(E(f_1), \dots, E(f_n)))))
\end{align}
where $E(\cdot)$ denotes the embedding of a face. 

\textbf{Overall Aggregation.} 
The above aggregation methods are applied sequentially, with the output of each step serving as the input for the next. This process ensures that each face embedding preserves its individual geometric and topological properties while incorporating contextual information from its neighboring vertices, edges, loops, and faces. The resulting face embeddings are then used as tokens in Transformers, enabling effective global feature learning across the entire B-rep model.

\subsection{Geometric Embedding of B-rep Models}
\label{sec:geoemtric-embedding}

\subsubsection{Vertex Embedding}
The vertex encoder takes as input the vertex's 3D coordinates and transforms them into high-dimensional vectors in the latent space. This transformation enables the vertex embeddings to be aligned with the edge embeddings in their respective dimensions. Specifically, the encoder, composed of 2 MLP layers with size 64, takes the 3D coordinates of the vertex as input and maps it to a latent space of dimension 64.

\subsubsection{Edge Embedding}
An edge is represented as a curve segment, which, in its most general form, is defined by a B-spline curve along with its start and end parameters. Directly inputting a B-spline curve into the neural network is challenging due to the varying number of control points, knots, and start/end parameters. To address this, existing methods discretize B-spline curves into fixed-size sampled points or polylines. However, this approach suffers: no universal discretization size suits all cases, and application-specific parameter tuning is required. These limitations highlight the need for a more flexible method that operates entirely in the continuous domain, eliminating discretization while ensuring a fixed-dimensional representation.

\textbf{Decomposing B-spline Curves into Bézier Curves.} 
Inspired by the geometric equivalence between B-spline and Bézier curves, we decompose a B-spline curve into a sequence of connected Bézier curves. A Bézier curve is a special case of a B-spline curve with no knots and a fixed number of control points when the degree is specified. Thus, it is a simpler curve formulation that can be effectively processed by neural networks. The decomposition is as follows.

A B-spline curve of degree $p$ is defined as:
\begin{equation}
\label{eq:bspline_curve}
\mathbf{C}(t) = \sum_{i=0}^{n} N_{i,p}(t) \mathbf{P}_{i},
\end{equation}  
where $t$ is the curve parameter, $N_{i,p}(t)$ are the B-spline basis functions of degree $p$, and $\mathbf{P}_{i}$ are the control points. $N_{i,p}(t)$ is defined recursively using the Cox-de Boor recursion:
\begin{equation}
    N_{i,0}(t)= 
    \begin{cases}
        1, & u_i \leq t < u_{i+1}, \\
        0, & \text{otherwise}.
    \end{cases}
\end{equation}
\begin{equation}
    N_{i,p}(t) = \frac{t - u_i}{u_{i+p} - u_i} N_{i,p-1}(t) + \frac{u_{i+p+1} - t}{u_{i+p+1} - u_{i+1}} N_{i+1,p-1}(t),
\end{equation}
where the knot vector $U = \{u_0, u_1, \cdots, u_{n+p+1}\}$ is a non-decreasing sequence of parameter values.

A Bézier curve of degree $ n $ is defined as:
\begin{equation}
    {\mathbf{B}(t) = \sum_{i=0}^{n} B_{i,n}(t) \mathbf{P}_i, \quad t \in [0,1]}
\end{equation}
where $ \mathbf{P}_i $ are the control points, and $ B_{i,n}(t) = \binom{n}{i} (1 - t)^{n-i} t^i $ are the Bernstein basis polynomials of degree $ n $.

\begin{figure}[t]
    \centering
    \includegraphics[width=\linewidth]{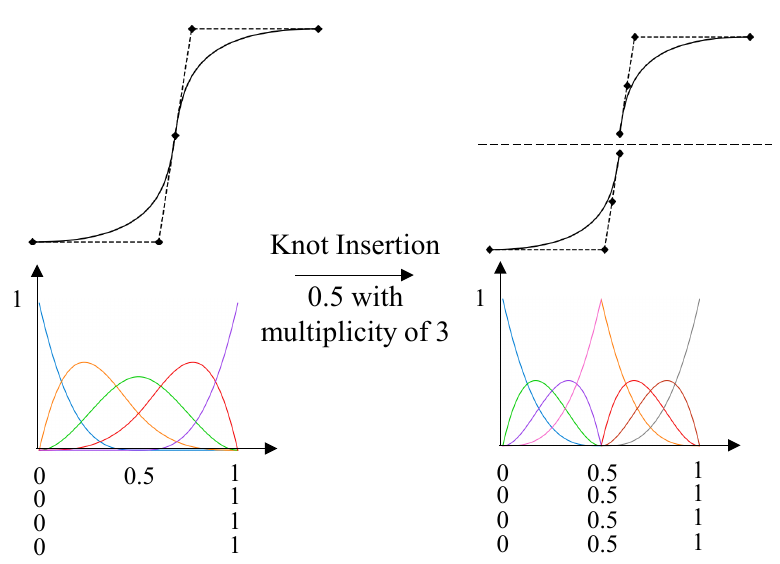}    
    \caption{Conversion of a B-spline curve into a sequence of Bézier curves with knot insertion.}
\label{fig:bspline2bezier-curve}
\end{figure}

To convert a B-spline curve into Bézier form, we employ the knot insertion technique, a process that increases the number of knots without altering the curve’s shape, as illustrated in Fig.~\ref{fig:bspline2bezier-curve}. Specifically, Boehm’s algorithm~\cite{piegl2012nurbs} is applied to insert knots at all interior breakpoints. Given a B-spline curve of degree $ p $ with control points $ \{ \mathbf{P}_i \} $ and a non-decreasing knot vector $ U = \{ u_0, u_1, \dots, u_m \} $, inserting a knot $ u \in [u_i, u_{i+1}] $ introduces a new control point $ \mathbf{P}_i' $ computed as a weighted average of existing control points:
\begin{equation}\label{eq:knot-insertion1}
    \mathbf{P}_i' = \alpha_i \mathbf{P}_{i-1} + (1 - \alpha_i) \mathbf{P}_i,
\end{equation}
where the blending factor $ \alpha_i $ is given by:
\begin{equation}
    \alpha_i = \frac{u - u_i}{u_{i+p} - u_i}, \quad u_i \leq u \leq u_{i+p}.
\end{equation}
By recursively applying this knot insertion until each segment is defined by exactly $ n+1 $ control points, where $ n $ is the degree of the curve, the B-spline curve is transformed into a sequence of Bézier segments, each spanning a single interval of the refined knot vector.

The above process allows us to represent the original B-spline curve entirely in Bézier form, facilitating subsequent processing steps while maintaining its geometric properties. In this work, we consistently use cubic Bézier curves for all cases, i.e., set the hyperparameter $n=3$.

\textbf{Reassembling Bézier Embeddings into B-spline Embeddings.}
After decomposing a B-spline curve into a series of Bézier segments, we employ an encoder network to learn the latent embeddings for each Bézier segment. The encoder consists of 6 MLP layers and takes as input the four control points of the (cubic) Bézier curve, as well as the tangent at the middle point of the curve, and then maps them to a latent space with dimension 64.

Given a sequence of Bézier curve embeddings $E(c_i)$ for $i = 0, 1, 2, \dots, n$, we aggregate these embeddings to obtain an embedding for the original B-spline curve using a Transformer encoder. Specifically, the embedding $E(e)$ is computed as:
\begin{equation}
    E(e) = MeanPooling\left(\text{Encoder}\left(E_{pos}(c_1),E_{pos}(c_2), \dots, E_{pos}(c_n)\right)\right) ,
\end{equation}
where $E_{\text{pos}}(c_i)$ denotes the $i$-th curve embedding added with cosine positional embedding. The resulting edge embedding $E(e)$ is extracted from the transformer's output at the first position, corresponding to $E'$. The encoder has 2 attention layers, 4 attention heads, and a hidden dimension of 512.

\subsubsection{Face Embedding}
\label{sec:face-embedding}
Similar to edge embedding, we decompose each B-spline surface into a collection of connected triangular Bézier patches, known as Bézier triangles. While rectangular Bézier patches (Bézier rectangles) are more commonly used, they struggle to match the shape of trimmed B-spline surfaces near their boundaries. In contrast, Bézier triangles offer greater flexibility while remaining continuous, knot-free, and having a fixed number of control points when the degree is specified, making them well-suited for deep learning.

\begin{figure}[t]
    \centering
    \includegraphics[width=\linewidth]{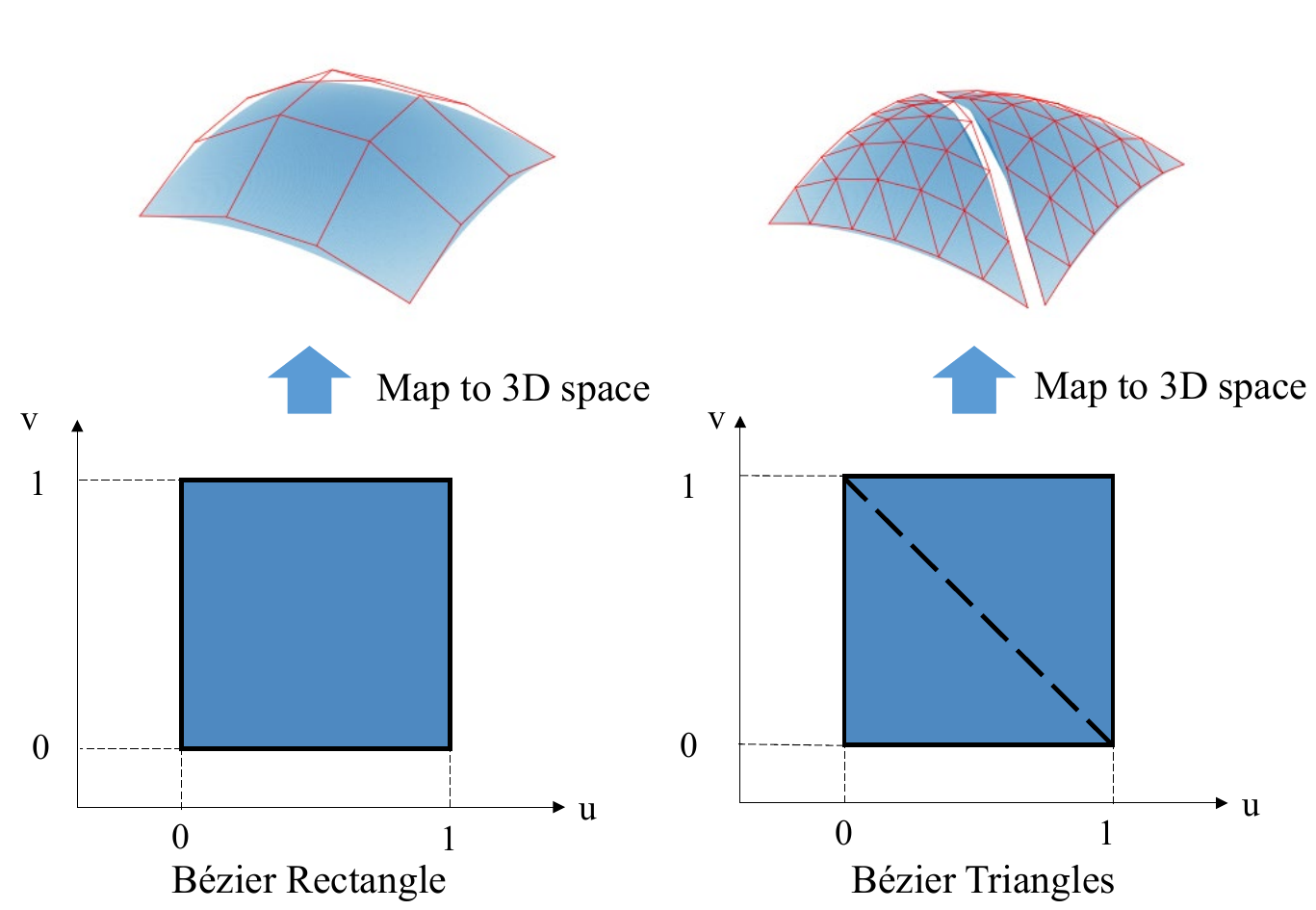}
    \caption{Convert Bézier rectangles into two Bézier triangles.}
    \label{fig:rec2tri}
\end{figure}

Our decomposition follows a two-step conversion. First, for regular (untrimmed) B-spline surfaces, we apply knot insertion to transform the surface into a grid of connected Bézier rectangles. Then, leveraging the property that a Bézier rectangle can be split into two Bézier triangles along its diagonal without altering the surface geometry~\cite{goldman1987conversion}, we further convert these Bézier rectangles into Bézier triangles (as shown in Fig.~\ref{fig:rec2tri}), resulting in a structured representation ideal for deep learning.

\begin{figure}[t]
    \centering
    \includegraphics[width=\linewidth]{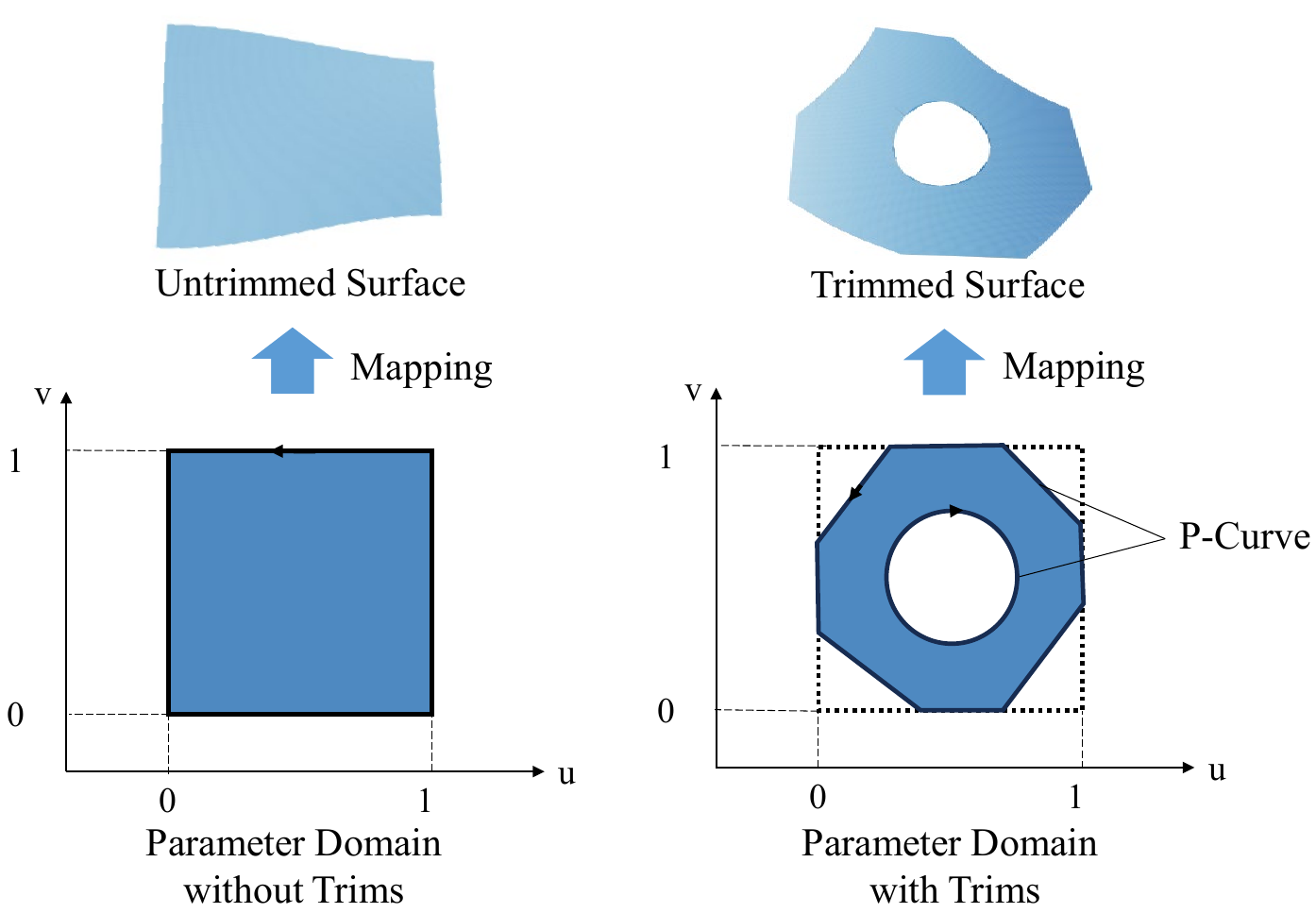}    
    \caption{B-spline surfaces: a complete surface and its rectangular parameter domain (left); a trimmed surface and its corresponding parameter domain (right).}
\label{fig:bspline-surface}
\end{figure}

For trimmed B-spline surfaces, we extend the above approach by segmenting the trimmed domain and embedding it within the Bézier rectangle representation using knot insertion. If a Bézier rectangle lies entirely within the trimmed parametric domain, the previous rectangle-to-triangle conversion applies directly. If a Bézier rectangle intersects the trimming curves in the parameter domain (p-curves, as illustrated in Fig.~\ref{fig:bspline-surface}), we perform the diagonal split as well, but with an additional step to refine the portion inside the trimmed region to ensure it aligns with the actual surface shape. This ensures that the final representation accurately preserves the true geometry of the trimmed surface while maintaining a structured Bézier triangle format. The following content provides a detailed explanation of this process.

\textbf{Surfaces and Their Trimming.} 
A B-spline surface $ \mathbf{S}(u, v) $ is defined as the tensor product of Eq.~\ref{eq:bspline_curve}:
\begin{equation}
    \mathbf{S}(u, v) = \sum_{i=0}^{n} \sum_{j=0}^{m} N_{i,p}(u) N_{j,q}(v) \mathbf{P}_{i,j}, \quad u, v \in [0,1],
\end{equation}
where $ \mathbf{P}_{i,j} $ are the control points, and $ N_{i,p}(u) $ and $ N_{j,q}(v) $ are the B-spline basis functions of degrees $ p $ and $ q $, respectively. The basis functions follow the Cox-de Boor recursion formula, similar to the curve case.

A trimmed surface consists of an underlying B-spline surface $ \mathbf{S}(u, v) $ and a set of p-curves, which are typically loops of B-spline curves defined in the parametric $(u,v)$ domain, as shown in Fig.~\ref{fig:bspline-surface}. These trimming curves outline the valid region of the surface while discarding the exterior portions. 

\textbf{Converting Untrimmed Surfaces to Bézier Triangles.}
The conversion process extends the 1D decomposition of B-spline curves into Bézier curves to 2D. First, we apply Eq.~\eqref{eq:knot-insertion1} along the $u$-direction to refine the surface in $u$-parameter space, followed by the same refinement along the $v$-direction. This two-step knot insertion increases the number of control points, forming a structured grid of Bézier rectangles that represent the same shape as the B-spline surface.

Once the B-spline surface is expressed as a grid of Bézier rectangles, each rectangle is split into two Bézier triangles along its diagonal by leveraging the method presented in~\cite{goldman1987conversion}. Let a Bézier rectangle of degree $ (m, n) $ be denoted by the following tensor-product form:
\begin{equation}
    \mathbf{S}(u, v) = \sum_{i=0}^{m} \sum_{j=0}^{n} B_{i}^{m}(u) B_{j}^{n}(v) \mathbf{P}_{i,j}, \quad 0 \leq u, v \leq 1,
\end{equation}
where $ B_{i}^{m}(u) $ and $ B_{j}^{n}(v) $ are Bernstein basis functions. The split Bézier triangle (the bottom-left one in Fig.~\ref{fig:rec2tri}) takes the following form:
\begin{equation}\label{eq:bezier-triangle-conversion}
    \mathbf{T}_1(s, t) = \sum_{a,b\geq 0}^{a+b \leq m+n}  B_{a,b}^{m+n}(s,t)  \mathbf{V}_{a,b}, \quad 0 \leq s, t, s+t \leq 1.
\end{equation}  
where the control points $\mathbf{V}_{a,b}$ are:
\begin{equation}\label{eq:bezier-triangle-control-points}
    \mathbf{V}_{a,b} = \frac{\sum_{i=0}^{a} \sum_{j=\max\{0,\,b-m+i\}}^{\min\{b,\,n-a+i\}}  \binom{a}{i} \binom{b}{j} \binom{m+n-a-b}{m-a-i}}{\binom{m+n}{n}} \mathbf{P}_{i,j},
\end{equation}  
and the bivariate Bernstein basis $B_{i,j}^{m+n}(s,t)$ is:
\begin{align}
    B_{a,b}^{m+n}(s,t) &= \binom{m+n}{a,b} s^a t^b (1-s-t)^{m+n-a-b} \\
    \binom{m+n}{a,b} &= \frac{(m+n)!}{a!b!(m+n-a-b)!}
\end{align}
Similarly, the second Bézier triangle (the top-right one) is derived by applying the same transformation symmetrically. Note that the derivation of Eq.~\ref{eq:bezier-triangle-conversion} presented in~\cite{goldman1987conversion} is complex. We have provided a simplified version in~\ref{app:B}. In this work, we consistently use rectangular Bézier patches with degrees $m=3$ and $n=3$.

\begin{figure}[t]
    \centering
    \includegraphics[width=\linewidth]{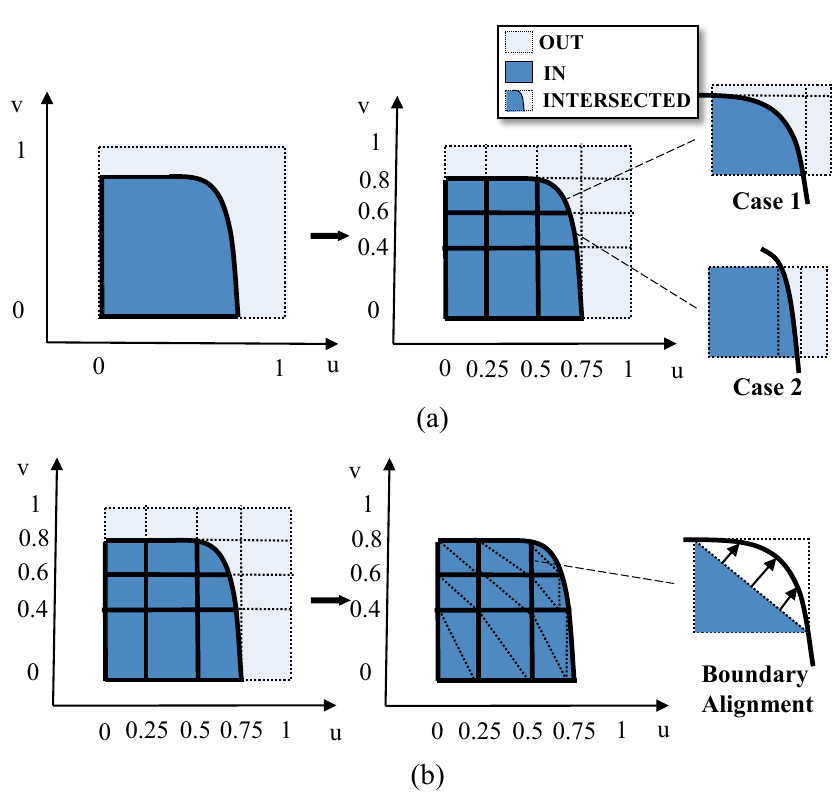}    
    \caption{Conversion of a trimmed B-spline surface into triangular Bézier patches: (a) Convert B-spline surface to Bézier rectangles; (b) Convert Bézier rectangles To Bézier triangles.}    
    \label{fig:bspline2bezier_trim}
\end{figure}

\textbf{Converting Trimmed Surfaces to Bézier Triangles.}  
The conversion here follows a similar workflow to that of untrimmed surfaces, with an additional step to handle Bézier rectangles intersecting p-curves, as shown in Fig.~\ref{fig:bspline2bezier_trim}a. Directly applying a diagonal split to such incomplete Bézier rectangles does not guarantee alignment with the actual surface shape along the surface boundary. To address this, we first subdivide the Bézier rectangle until the p-curve passes through a pair of its diagonal points (Fig.~\ref{fig:bspline2bezier_trim}b). This ensures a precise classification of the intersection region, allowing for a clean diagonal split.

Once subdivided, we apply the standard diagonal split and identify the Bézier triangle within the trimming domain. To ensure geometric alignment with the original surface along the trimming boundary, we refine its control points using the following least-squares optimization:
\begin{equation}
   \min_{\mathbf{V}_{i,j}} \sum_{k=0}^{N}  \left\| \mathbf{P}_{k} - \sum_{i,j\geq 0}^{i+j \leq m+n}  B_{i,j}^{m+n}(u_k,v_k)  \mathbf{V}_{i,j} \right\|^2
   + \lambda \sum_{i,j\geq 0}^{i+j \leq m+n} \left\| \mathbf{V}_{i,j} - \mathbf{V}_{i,j}^{\text{orig}} \right\|^2
\end{equation}
where $\mathbf{P}_k$ are sampled surface points along the trimming boundary, $\mathbf{V}_{i,j}^{\text{orig}}$ are the control points obtained from the initial rectangle-to-triangle conversion (as in Eq.~\ref{eq:bezier-triangle-control-points}), and $\lambda$ is a weighting parameter controlling the balance between surface fitting and shape preservation. The first term ensures the Bézier triangle closely aligns with the surface along the trimming boundary, while the second term prevents excessive deviation from its original shape. This optimization problem is a standard least-squares problem, which reduces to solving a linear system.
Finally, the resulting Bézier triangles not only ensure smooth transitions with existing triangles but also maintain a well-structured Bézier triangle framework.

\textbf{The Order in Handling Bézier Triangles.}
The triangular Bézier patches are processed following a z-order curve traversal, which is the standard way to handle tree-like input data in deep learning, typically~\cite{wang2023octformer}.  Specifically, we subdivide the surface by recursively dividing each patch into four smaller patches, forming a quadtree structure where each leaf node represents a Bézier patch.  These patches are then sorted using shuffled keys. Denote the integer coordinates of a patch as $(x,y)$ and the i-th bit of each coordinate's bits as $x_i$ and $y_i$, the shuffled keys in binary expression are detailed as follows:
\begin{equation}
    {Key(x,y)= x_1 y_1 x_2 y_2 \dots x_d y_d}
\end{equation}
where d is the depth of the patch.  When visiting each rectangular Bézier patch in this order, a z-order curve is formed.  Each rectangular patch contains one or two triangular Bézier patches, which are visited in a defined order within that patch.

\textbf{Bézier Triangle Embedding.}
Each Bézier triangle can be represented by its control points. For a Bézier triangle of degree $2p$, the number of control points is $(2p+1)(2p+2)/2$, with each control point consisting of 3D-coordinates and weight, forming a tensor of size $(2p+1)(p+1) \times 4$. We also include surface normals at Bézier triangle center points as additional input features. These normal vectors are concatenated with the tensor of control points. The new tensor is then fed into a 6-layer MLP encoder, which maps its input to a 64-dimensional latent space.

\textbf{Reassemble Triangular Bézier Embeddings into B-spline Surface Embeddings.} 
Given a sequence of triangular Bézier embeddings $E(t_i)$ for $i = 0, 1, 2, \dots, n$, we aggregate these embeddings to obtain an embedding for the original B-spline surface using a Transformer encoder. Specifically, the embedding $E(f)$ is computed as:
\begin{equation}
    E(f) = MeanPooling\left(\text{Encoder}\left(E_{pos}(t_1),E_{pos}(t_2), \dots, E_{pos}(t_n)\right)\right) ,
\end{equation}
where $E_{\text{pos}}(t_i)$ denotes the $i$-th  triangular Bézier embedding added with cosine positional embedding. The resulting face embedding $E(f)$ is extracted from the transformer's output at the first position, corresponding to $E'$. The encoder has 2 attention layers, 8 attention heads, and a hidden dimension of 512.

\textit{Remark:} The B-spline surfaces to Bézier triangles decomposition process in BRT can also introduce errors when surfaces exceed the fixed Bézier degree used. However, these errors can be eliminated by selecting a sufficiently high Bézier degree, e.g., the maximum degree in the dataset.

\subsection{Training BRT}
\label{sec:train_brt}
By integrating the preceding geometric and topological embedding modules, we obtain a B-rep tokenization network, which, when paired with a Transformer encoder (without the positional encoding component), forms the BRT network. We train the BRT on each domain-specific task, such as part classification and feature recognition. A masking technique is also introduced during training to enhance BRT's robustness and generalizability.

\textbf{Training by Domain-Specific Tasks.}
We train the BRT on key B-rep modeling tasks like part classification and feature recognition, which have applications like CAD model searching and CAD/CAM integration~\cite{jayaraman2021uv,wu2024aagnet}. Labeled data for this training comes from a proprietary dataset TMCAD collected by our group (will be released upon publication), as well as publicly available datasets: FebWave~\cite{angrish2019fabsearch}, SolidLetters~\cite{jayaraman2021uv}, MFCAD++~\cite{colligan2022hierarchical}, and Fusion 360 Gallery~\cite{lambourne2021brepnet}.

\textbf{Training by Masking.}
To enhance the robustness and generalizability of BRT, we apply a masking technique~\cite{he2022masked} that randomly drops a subset (e.g., 25\% or even 50\%) of Bézier triangles during training and then forces BRT to continue to generate the same correct outputs despite the missing data. This encourages BRT to learn the structural and contextual relationships within B-rep models, rather than memorizing exact input values, which can lead to overfitting. 

\section{Results}
\label{sec:results}
In this section, experiments have been conducted to evaluate BRT on classification and segmentation tasks on various datasets. The BRT is trained using PyTorch and runs on an NVIDIA GeForce RTX 4090D GPU. The training uses the ADAM optimizer~\cite{kingma2014adam} with a learning rate of 0.0001. The size of each batch is 16, and we perform the BRT training with only 350 epochs. The part classification, machining feature recognition, and model segmentation experiments are performed, which are detailed in Sec.~\ref{sec:exp_classification}, Sec.~\ref{sec:exp_segmentation}, and Sec.~\ref{sec:exp_segmentation_operation}, respectively. The code for BRT is available at: https://github.com/Qiang-Zou/BRT

\subsection{Datasets}
\textbf{TMCAD Dataset}. The TMCAD (Truly Mechanical CAD) dataset comprises more than 10,000 real-world mechanical design models collected from the Internet. The data is split into 10 classes: \textit{Bearing}, \textit{Bolt}, \textit{Bracket},  \textit{Coupling},  \textit{Flange},  \textit{Gear}, \textit{Nut},   \textit{Pulley},   \textit{Screw},  and \textit{Shaft}. They are all stored in STEP file format. For more details and sample models, refer to \ref{app:D}. Our construction of TMCAD is an ongoing effort, and the number of CAD models continues to grow. The dataset is available at: https://github.com/Qiang-Zou/BRT.

\textbf{FebWave Dataset}. The FebWave dataset comprises 4,572 labeled 3D shapes categorized into 45 classes, such as brackets, gears, and o-rings. Note that the overall FebWave dataset consists of more than 100,000 models, but only a very small portion of it has labels. For more details, refer to the paper~\cite{angrish2019fabsearch}.

\textbf{SolidLetters Dataset}. The SolidLetters dataset comprises 96k 3D shapes generated by randomly extruding and filleting the 26 alphabet letters. These shapes are organized into 2,002 style categories based on different fonts. This dataset is quite artificial, but its labeling quality is high. For more details, refer to the paper~\cite{jayaraman2021uv}.

\textbf{MFCAD++ Dataset}. This dataset contains 59,655 CAD models with 24 types of machining features. Each CAD model includes 3 to 10 machining features. For more details, refer to the paper~\cite{colligan2022hierarchical}.

\textbf{Fusion 360 Gallery Dataset}. The Fusion 360 Gallery dataset has 35,858 3D models sourced from Autodesk's Fusion 360 platform. It provides segmentation labels, which are done by assigning one of eight modeling operation categories to each face: ExtrudeSide, ExtrudeEnd, CutSide, CutEnd, Fillet, Chamfer, RevolveSide, and RevolveEnd. For more details, refer to the paper~\cite{lambourne2021brepnet}.

\subsection{Comparisons}
This section presents comparison results based on the datasets stated above. For all datasets, we employ a split of 70\% for training, 15\% for validation, and 15\% for testing. The baseline models used for comparison include UV-Net~\cite{jayaraman2021uv}, BrepNet~\cite{lambourne2021brepnet}, and AAGNet~\cite{wu2024aagnet}.  All models were implemented using their official open-source code with default settings (see \ref{app:C} for details), which we assume have been fine-tuned by the original authors for optimal performance. Masking was not applied to these models for two main reasons: (1) they inherently incorporate locality bias through convolutional kernels or message-passing mechanisms, making explicit masking less essential; and (2) incorporating masking into CNNs or GNNs---the backbone architectures of UV-Net, AAGNet, and BrepNet---is nontrivial and fundamentally different from how transformers implement it~\cite{he2022masked}.

\subsubsection{3D Shape Classification}
\label{sec:exp_classification}
We first evaluate our method on the task of 3D shape classification. 
For comparison, we train UV-Net~\cite{jayaraman2021uv} using 10x10 grids, and AAGNet~\cite{wu2024aagnet} with all attributes enabled. Since AAGNet is originally designed for segmentation tasks, we adapt it for classification by utilizing the global features produced by its graph encoder as the B-rep model's global features and adding a classification head composed of two MLP layers to generate the classification logits. We exclude BrepNet~\cite{lambourne2021brepnet} from the comparison because its network architecture is not tailored for classification tasks.

All models are trained for up to 350 epochs using cross-entropy loss and the Adam optimizer with $\beta_1 = 0.001$ and $\beta_2 = 0.999$. As shown in Table~\ref{tab:classification_results}, our method achieves the highest classification accuracy across all datasets. This highlights the effectiveness of our continuous geometric and hierarchical topological embeddings, enabling our approach to outperform existing methods. 

\begin{figure}[htbp]
    \centering
    \includegraphics[width=\linewidth]{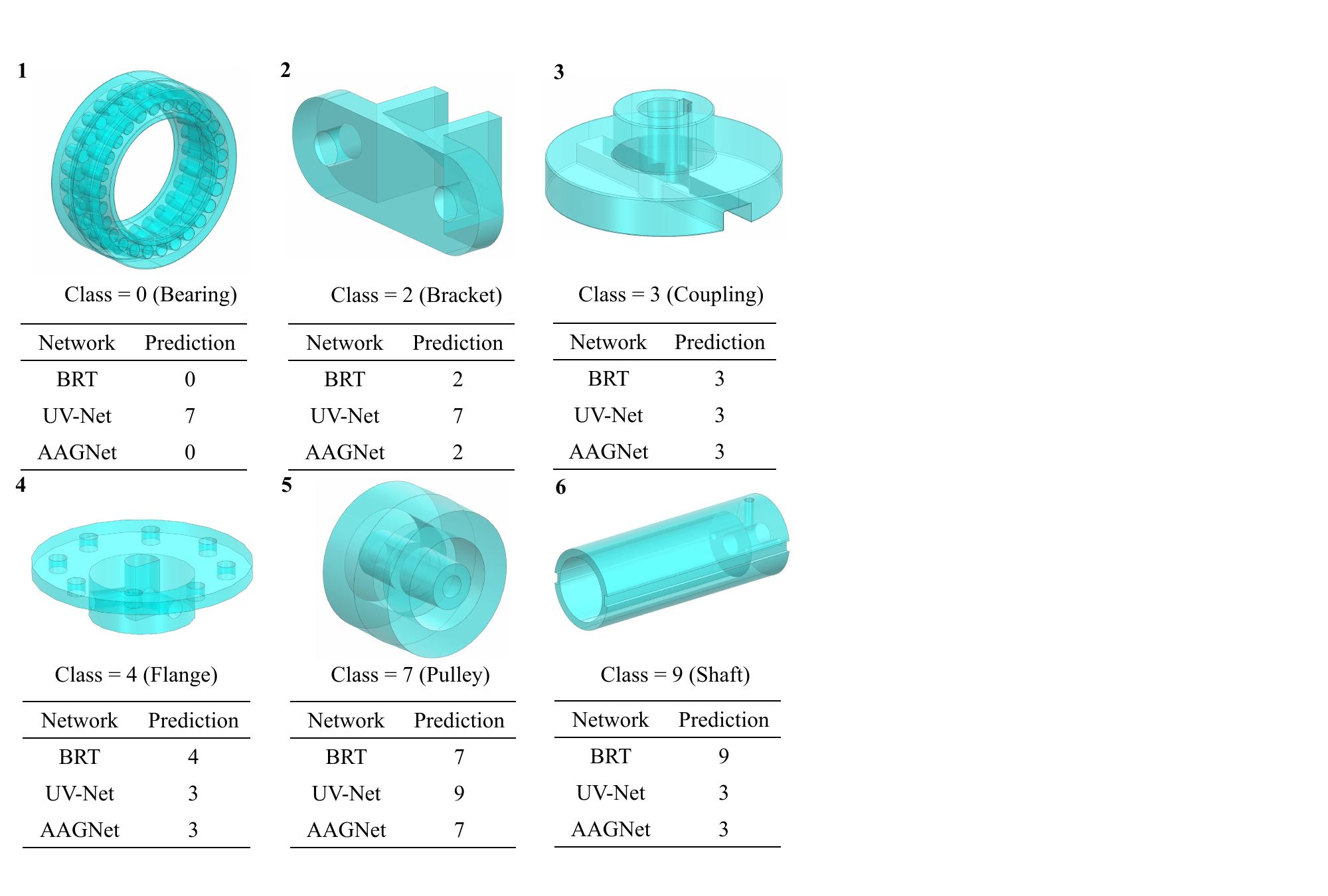}
    \caption{Examples of classification results on TMCAD.}
    \label{fig:example_classification}
\end{figure}

Fig.~\ref{fig:example_classification} illustrates some classification results on the test set of the TMCAD dataset compared with two other methods. BRT successfully predicts all these samples, compared to two other methods.

\begin{table}[htbp]
    \centering
    \caption{Classification performance comparison across datasets.}
    \label{tab:classification_results}
    \begin{tabular}{lclc}  
        \toprule
        Dataset & Network & Input Geometry & Accuracy (\%) \\
        \midrule
        \multirow{3}{*}{TMCAD}
            & BRT (Ours) & Bézier patches & \textbf{83.45} \\
            & UV-Net    & Grids  & 81.31         \\
            & AAGNet    & Grids+ Attributes   & 74.72         \\
        \addlinespace
        \multirow{3}{*}{FebWave} 
            & BRT (Ours) & Bézier patches & \textbf{98.95} \\
            & UV-Net    & Grids   & 93.49         \\
            & AAGNet    & Grids+ Attributes  & 96.33         \\
        \addlinespace
        \multirow{3}{*}{SolidLetters} 
            & BRT (Ours) & Bézier patches & \textbf{97.35} \\
            & UV-Net   & Grids   & 95.82         \\
            & AAGNet    & Grids + Attributes   & 96.99         \\
        \bottomrule
    \end{tabular}
\end{table}

\subsubsection{Machining Feature Recognition}
\label{sec:exp_segmentation}

In the machining feature recognition tasks, we compared BRT with UV-Net~\cite{jayaraman2021uv}, AAGNet~\cite{wu2024aagnet}, and BrepNet~\cite{lambourne2021brepnet} using the MFCAD++ dataset~\cite{colligan2022hierarchical}. We train UV-Net and AAGNet using network configurations similar to those employed in the classification experiments, and we train BrepNet with the winged edge kernel.

The evaluation metrics are defined as follows:
\begin{equation}
    \text{Accuracy} = \frac{\text{Number of correctly classified faces}}{\text{Total number of faces}}
\end{equation}
\begin{equation}
    \text{IoU} = \frac{1}{C}\sum_{c=1}^{C}\frac{A_c \cap B_c}{A_c \cup B_c}
\end{equation}
where $C$ is the number of classes, $A_c$ is the set of faces that belong to class $c$ in the ground truth label, and $B_c$ is the set of faces that belong to class $c$ in the predicted label. The IoU is so defined to measure the overlap between the predicted class of each face and the ground truth. These metrics are also used in Sec.~\ref{sec:exp_segmentation_operation}.

\begin{figure*}[htbp]
    \centering
    \includegraphics[width=\linewidth]{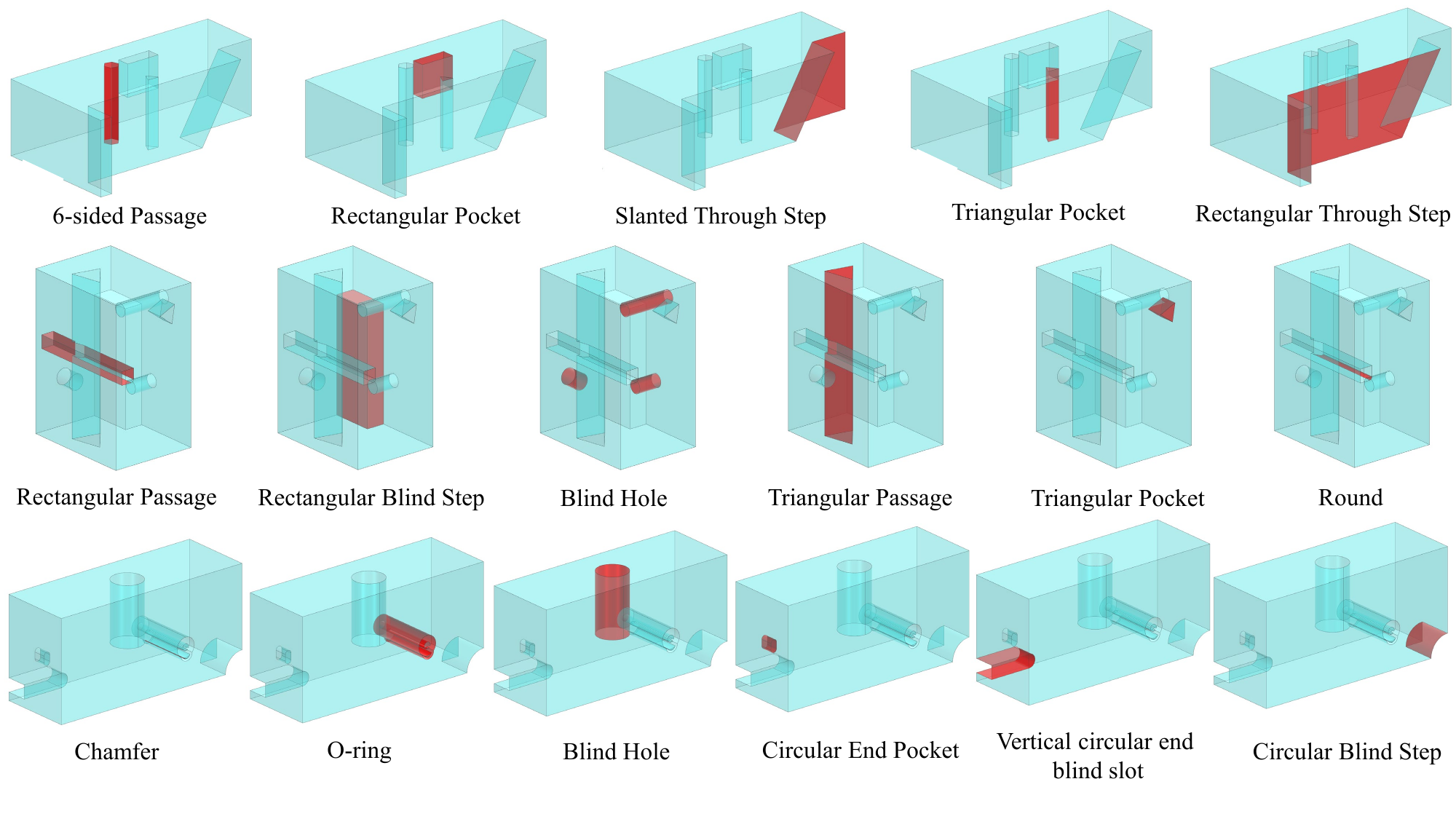}
    \caption{Examples of feature recognition on MFCAD++.}    \label{fig:example_segmentation}
\end{figure*}

Table~\ref{tab:segmentation_results} presents the feature recognition accuracy and mean Intersection over Union (IoU) for each network on MFCAD++. Our method outperforms both UV-Net and BrepNet, as demonstrated by some of our feature recognition results in Fig.~\ref{fig:example_segmentation}. However, AAGNet achieves slightly higher accuracy and IoU than our approach. This is because AAGNet effectively handles planar faces, which constitute the majority of faces in the MFCAD++ dataset. In contrast, our continuous geometry representation does not leverage its advantages on planar faces as effectively within this dataset, unlike in datasets with more complex geometries such as TMCAD and SolidLetters.

\begin{table}[h!]
    \centering
    \caption{Feature recognition performance comparison on MFCAD++.}
    \label{tab:segmentation_results}
    \begin{tabular}{ccc}
        \toprule
        Network & Accuracy (\%) & IoU (\%) \\
        \midrule

                           BRT (Ours)          & 99.27 & 97.98  \\
                           UV-Net              & 99.02 & 96.85   \\
                           AAGNet & 99.29 & 98.64  \\
                           BrepNet          & 99.22 & 97.75  \\
        \bottomrule
    \end{tabular}
\end{table}

\subsubsection{Segmentation on CAD Modeling Operations}
\label{sec:exp_segmentation_operation}
This subsection presents experimental results on the Fusion 360 Gallery Dataset~\cite{lambourne2021brepnet} to further evaluate the performance of BRT on the segmentation task involving more complex B-rep models.

Table~\ref{tab:segmentation_results2} shows the face accuracy and mean Intersection over Union (IoU) for each network on the Fusion 360 Gallery dataset. Our method outperforms all other approaches in terms of face accuracy and IoU. Notably, the Fusion 360 Gallery dataset, derived from real-world data, exhibits higher structural and geometric complexity compared to MFCAD++. Consequently, the accuracy and IoU of all methods are lower on this dataset compared to MFCAD++. The class-wise performance comparisons presented in Table~\ref{tab:segmentation_results2_classwise} further demonstrate the advantage of our method across almost all classes.

\begin{table}[h!]
    \centering
    \caption{Segmentation performance comparisons on Fusion 360 Gallery.}
    \label{tab:segmentation_results2}
    \begin{tabular}{ccc}
        \toprule
         Network & Accuracy (\%) & IoU (\%)  \\
        \midrule 
                           BRT (Ours)          & \textbf{94.48}   & \textbf{79.23} \\
                           UV-Net              & 89.03 & 66.47 \\
                           AAGNet & 82.45 & 75.53 \\
                           BrepNet          & 90.19 & 68.92  \\
        \bottomrule
    \end{tabular}
\end{table}

\begin{table*}[h]
\centering
\caption{Class-wise segmentation performance comparisons on Fusion 360 Gallery. Note that accuracy is calculated per class, so the average does not match the overall accuracy.}
\label{tab:segmentation_results2_classwise}
\begin{tabular}{l|cc|cc|cc|cc}
\toprule
\textbf{Class} & \multicolumn{2}{c|}{\textbf{BRT (Ours)}} & \multicolumn{2}{c|}{\textbf{UV-Net}} & \multicolumn{2}{c|}{\textbf{AAGNet}} & \multicolumn{2}{c}{\textbf{BrepNet}} \\
 & IoU (\%) & Acc (\%) & IoU (\%) & Acc (\%) & IoU (\%) & Acc (\%) & IoU (\%) & Acc (\%) \\
\midrule
ExtrudeSide  & \textbf{93.43} & \textbf{97.12} & 84.92 & 93.92 & 88.69 & 94.60 & 90.03 & 96.48 \\
ExtrudeEnd   & \textbf{91.52} & \textbf{96.54} & 85.69 & 94.02 & 89.38 & 94.21 & 87.47 & 93.93 \\
CutSide      & \textbf{80.63} & \textbf{87.47} & 61.24 & 72.26 & 67.34 & 76.87 & 76.91 & 76.30 \\
CutEnd       & \textbf{63.94} & \textbf{76.75} & 56.84 & 67.61 & 63.91 & 76.72 & 63.55 & 68.80 \\
Fillet       & \textbf{89.47} & \textbf{93.69} & 79.01 & 87.99 & 88.87 & 89.81 & 82.89 & 84.53 \\
Chamfer      & \textbf{82.97} & \textbf{90.63} & 80.40 & 89.24 & 80.54 & 89.98 & 76.71 & 72.43 \\
RevolveSide  & \textbf{84.08} & \textbf{90.16} & 69.98 & 73.92 & 78.55 & 87.86 & 67.48 & 83.21 \\
RevolveEnd   & \textbf{47.83} & 51.56 & 13.67 & 12.31 & 46.98 & \textbf{54.05} & 6.25  & 50.00 \\
\bottomrule
\end{tabular}
\end{table*}

\subsubsection{Discussions}
The comparisons were carried out using five datasets: MFCAD++, Fusion 360 Gallery, SolidLetters, FabWave, and TMCAD. Our method achieves SOTA accuracy on Fusion 360 Gallery, SolidLetters, FabWave, and TMCAD, with improvements of 4.29\%, 0.36\%, 2.62\%, and 5.16\%, respectively. Admittedly, the performance gains on SolidLetters is marginal. This is likely due to the plane-dominant nature of these datasets, which limits the advantage of our continuous geometry-based tokenization over discretization-based methods.  In contrast, our method excels with more complex, freeform geometries (such as FabWave and TMCAD) due to its ability to preserve B-spline surface shapes.  This suggests that our method’s strength lies in scenarios where fine-grained geometric detail matters.

On MFCAD++, our method ranks second, with AAGNet at 99.29\% and ours at 99.27\%. However, the performance gap among all methods on this dataset is not significant.  This is because MFCAD++ is composed largely of plane models, and this does not present sufficient geometric complexity to differentiate among current methods.  As a result, nearly all methods perform sufficiently well and similarly well.

That said, our method shows advantages on datasets with intricate geometries, where discretization struggles.  This implies the promise of continuous geometry-based learning methods, and further performance gains may be expected when larger and more diverse B-rep datasets become available.

\subsection{Ablation Study}
In this subsection, we examine the impact of various input features and network components on the machining feature recognition task using the MFCAD++ dataset. The study is structured as follows:
\begin{enumerate}[label=(\alph*)]
    \item \label{ournet}\textbf{BRT (Original):} The BRT network, incorporating all input features and modules.
    \item \textbf{w.o. Masking:} We removed the masking strategy described in Sec.~\ref{sec:train_brt}.
    \item \textbf{w.o. Trim:} A variant of BRT that disregards the trimming curves of each B-rep face and transforms the untrimmed surfaces into Bézier triangles, as outlined in Sec.~\ref{sec:face-embedding}.
    \item \textbf{GCN:} We replaced the original topological encoder with a Graph Convolutional Network (GCN) module.
    \item \textbf{UV-grids:} We replaced the Bézier triangles with $10 \times 10$ UV-grids.
    \item \textbf{Convolution:} We handle the Bézier triangles with GCN instead of transformer. Here, neighboring relationships of patches are treated as a graph.
\end{enumerate}

The ablation study results are presented in Table~\ref{tab:ablation_studies_input_features}. Our findings indicate that the masking strategy significantly enhances the performance of the BRT model. Additionally, the model utilizing trimmed surfaces performs slightly better than its non-trimmed counterpart. Moreover, compared to the GCN, which serves as a general-purpose module for graph-based data, our specialized topological encoder demonstrates superior performance on the B-rep MFCAD++ dataset. Further ablations demonstrate the Transformer-based face encoder’s effectiveness: replacing it with a convolutional network or substituting the entire encoder with a UV-grid surface encoder degrades performance. These experiments confirm the effectiveness of our proposed methods.

    \begin{table}[h!]
        \centering
        \caption{Ablation study with input features and components of BRT on the MFCAD++ for feature recognition task.}
        \label{tab:ablation_studies_input_features}
            \begin{tabular}{lc}
                \toprule
                Model & Accuracy (\%) \\
                \midrule
                BRT (Original)           & \textbf{99.27} \\
                \textbf{w.o.} Masking              & 94.79 \\
                \textbf{w.o.} Trim     & 99.01 \\
                GCN & 98.72 \\
                {UV-grids} & {98.14} \\
                {Convolution} & 97.78 \\
                \bottomrule
            \end{tabular}
    \end{table}

\section{Conclusion}
\label{sec:conclusion}
In this paper, we introduced BRT, a novel tokenization method for learning B-rep data with Transformers. Our approach involves converting faces into triangular Bézier patches and edges into Bézier curves, thereby preserving the continuous geometry in B-rep models. Additionally, we developed a topological embedding strategy to effectively represent the irregular structure of B-rep data. This approach demonstrated state-of-the-art performance in both classification and segmentation tasks across several B-rep datasets. Furthermore, we introduced a new B-rep dataset, TMCAD, which consists completely of real-world data, underscoring the practical applicability of our work.

Despite BRT's demonstrated effectiveness on some datasets, it still has certain limitations and, therefore, future research problems. During the conversion of trimmed B-spline surfaces into triangular Bézier patches, ensuring that the Bézier triangles on the boundaries precisely conform to the boundary curves can be challenging. Often, these boundary curves cannot be exactly represented as B-spline curves, leading to approximation errors. Fortunately, such an error is limited only to boundary curves and thus ignorable in most cases.

Another concern arises in that the Transformer is vulnerable to long sequences unseen during training.  The maximum number of curves that our model can process is 100.  This limit was chosen because, from our statistics, most (more than 99\%) of edges in practical B rep models decompose into fewer than 100 Bézier curve segments.  In practice, the curve sequences in our datasets have not exceeded this limit.  To further investigate this, we analyzed an additional 300 B-rep models sampled from grabcad.com (the largest online CAD model repository).  Fortunately, edges with more than 100 curve segments constituted less than 0.04\% among all the edges in these models.

Moreover, while transformers have demonstrated exceptional performance with large datasets, the current availability of B-rep data remains limited. With larger and more diverse datasets emerging, BRT is likely to unlock significant potential. Also, the transformer decoder could be utilized to autoregressively generate B-rep models, further enhancing the capabilities and applications of our approach.

\section*{Acknowledgements}
This work has been funded by the “Pioneer” and “Leading Goose” R\&D Program of Zhejiang Province (No. 2024C01103), the National Natural Science Foundation of China (No. 62102355), Natural Science Foundation of Zhejiang Province (No. LQ22F020012), and the Fundamental Research Funds for the Zhejiang Provincial Universities (No. 2023QZJH32).


 \bibliographystyle{elsarticle-num} 
 \bibliography{elsarticle-template-num}






\appendix

\section{Bézier rectangles to Bézier Triangles}
\label{app:B}

A Bézier rectangle of degree $ (m, n) $ in the parametric domain $ (u, v) $ is given by the tensor-product form:
\begin{equation}
    \mathbf{S}(u, v) = \sum_{i=0}^{m} \sum_{j=0}^{n} B_{i}^{m}(u) B_{j}^{n}(v) \mathbf{P}_{i,j}, \quad 0 \leq u, v \leq 1,
\end{equation}
where $ B_{i}^{m}(u) $ and $ B_{j}^{n}(v) $ are Bernstein basis polynomials.

To split the rectangle along the diagonal from $ (0,1) $ to $ (1,0) $ (Fig.~\ref{fig:rec2tri}), we introduce two sets of barycentric coordinates:
\begin{align}
    {s} &{= u}, & {t} &= {v}, & {s + t} &{\leq 1}, \\
    {s'} &= {1 - u,} & {t'} &{= 1 - v,} & {s' + t'} &{\leq 1.}
\end{align}

Rewriting the univariate Bernstein polynomials in terms of these new barycentric coordinates:
\begin{equation}
    B_{i}^{m}(u) = \sum_{h=0}^{m-i} B_{i,h}^{m}(s,t), \quad
    B_{j}^{n}(v) = \sum_{k=0}^{n-j} B_{k,j}^{n}(s,t),
\end{equation}
where the bivariate Bernstein basis is:
\begin{equation}
    B_{i,h}^{m}(s,t) = \binom{m}{i,h} s^i t^h (1-s-t)^{m-i-h}, \quad
    \binom{m}{i,h} = \frac{m!}{i!h!(m-i-h)!}.
\end{equation}

Thus, their product expands as:
\begin{equation}
    B_i^m(u) B_j^n(v) = \sum_{h=0}^{m-i} \sum_{k=0}^{n-j} B_{i,h}^{m}(s,t) B_{k,j}^{n}(s,t).
\end{equation}

Using the identity:
\begin{equation}
    B_{i,h}^{m}(s,t) B_{k,j}^{n}(s,t) = \frac{\binom{m}{i,h} \binom{n}{k,j} \binom{m+n}{i+k,h+j}}{\binom{m+n}{i+k,h+j}} B_{i+k,h+j}^{m+n}(s,t),
\end{equation}
we obtain:
\begin{equation}
    B_i^m(u) B_j^n(v) = \sum_{h=0}^{m-i} \sum_{k=0}^{n-j} \frac{\binom{i+k}{i} \binom{h+j}{j} \binom{m+n-k-i-h-j}{m-i-h}}{\binom{m+n}{n}} B_{i+k,h+j}^{m+n}(s,t).
\end{equation}

This means we can rewrite the Bézier rectangle $ S(u,v) $ as:
\begin{equation}
    \mathbf{S}(u, v) = \sum_{i=0}^{m} \sum_{j=0}^{n} \sum_{h=0}^{m-i} \sum_{k=0}^{n-j} \frac{\binom{i+k}{i} \binom{h+j}{j} \binom{m+n-k-i-h-j}{m-i-h}}{\binom{m+n}{n}} \mathbf{P}_{i,j} B_{i+k,h+j}^{m+n}(s,t).
\end{equation}

By defining new control points:
\begin{equation}
    \mathbf{V}_{a,b} = \frac{\sum_{i=0}^{a} \sum_{j=\max\{0,\,b-m+i\}}^{\min\{b,\,n-a+i\}}  \binom{a}{i} \binom{b}{j} \binom{m+n-a-b}{m-a-i}}{\binom{m+n}{n}} \mathbf{P}_{i,j},
\end{equation}  
we can rewrite $ \mathbf{S}(u,v) $ in terms of barycentric coordinates:
\begin{equation}
    \mathbf{S}(u, v) = \sum_{a,b\geq 0}^{a+b \leq m+n}  B_{a,b}^{m+n}(s,t) \mathbf{V}_{a,b}.
\end{equation}  
Thus, the resulting Bézier triangle (the bottom-left one) is: 
\begin{equation}
    \mathbf{T}_1(s, t) = \sum_{i,j\geq 0}^{i+j \leq m+n}  B_{i,j}^{m+n}(s,t)  \mathbf{V}_{i,j}, \quad 0 \leq s, t, s+t \leq 1.
\end{equation}  
Similarly, the second Bézier triangle (the top-right one) is derived by applying the same transformation symmetrically.

\section{Implementation details}
\label{app:C}
The sizes of all networks in this work are as follows.
\begin{itemize}
    \item BRT comprises primarily six attention layers, with an embedding size of 64 and a hidden size of 512, totaling 2.09M parameters.
    \item UV-Net comprises primarily 3 Conv1d layers and 3 Conv2d layers with an embedding size of 64 and hidden size of 64, resulting in 1.36M parameters.
    \item AAGNet comprises primarily 3 Conv2d layers, with an embedding size of 64 and hidden size of 256, totaling 0.38M parameters—the most lightweight among the compared methods.
    \item BrepNet comprises primarily 3 Conv1d layers and 3 Conv2d layers, with an embedding size of 64 and hidden size of 2100, totaling 1.84M parameters.
\end{itemize}

The hyperparameters used in processing the input representations for UV-Net, AAGNet, and BrepNet are given in Table~\ref{tab:network_configs}.

\begin{table}[h!]
    \centering
    \caption{The input representation parameters of UV-Net, AAGNet, and BrepNet.}
    \label{tab:network_configs}
    \small  
    \begin{tabular}{P{3.2cm}ccc}
        \toprule
        Parameters & UV-Net & AAGNet & BrepNet \\
        \midrule
        Edge Grid Dim & 6 & / & / \\
        Coedge Grid Dim & / & / & 12 \\
        Surface Grids Dim & 7 & 7 & 7 \\
        Surface Grids Size & 10x10 & 5x5 & 10x10 \\
        Edge Grids Size & 10 & / & / \\
        Coedge Grids Size & / & / & 10 \\
        Edge Embedding Dim & 64 & 64 & 64 \\
        Surface Embedding Dim & 64 & 64 & 64 \\
        Edge Attr.\ Dim & / & 12 & 10 \\
        Coedge Attr.\ Dim & / & / & 1 \\
        Surface Attr.\ Dim & / & 10 & 7 \\
        Attributes & / & all & all \\
        Kernel & / & / & Winged\newline Edge++ \\
        \bottomrule
    \end{tabular}
\end{table}

\section{TMCAD Dataset}
\label{app:D}
\begin{figure*}[htbp]
    \centering
    \includegraphics[width=.75\linewidth]{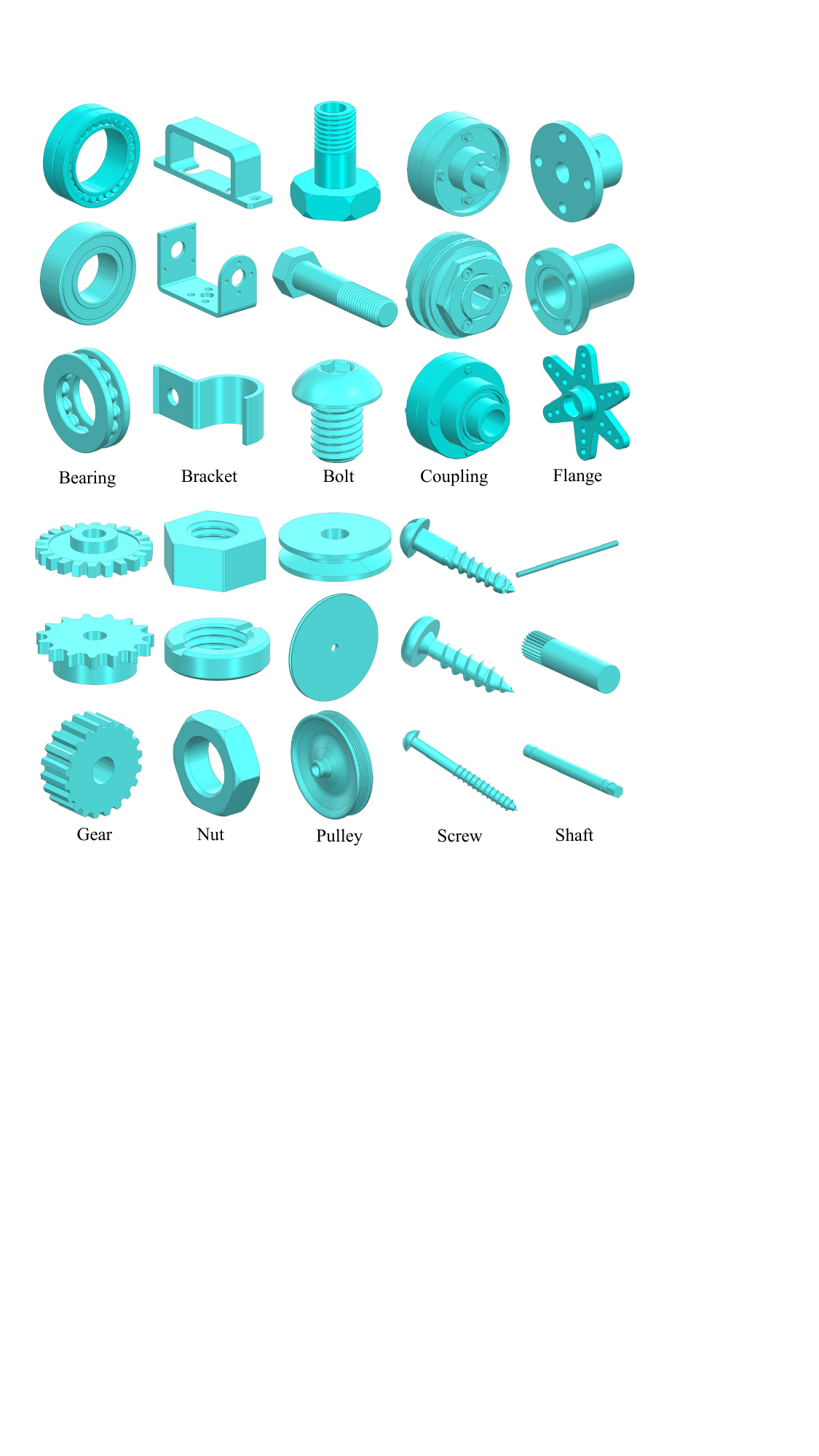}
    \caption{Model examples in the TMCAD dataset.}
    \label{fig:TMCAD}
\end{figure*}

\begin{figure*}[htbp]
    \centering
    \includegraphics[width=.75\linewidth]{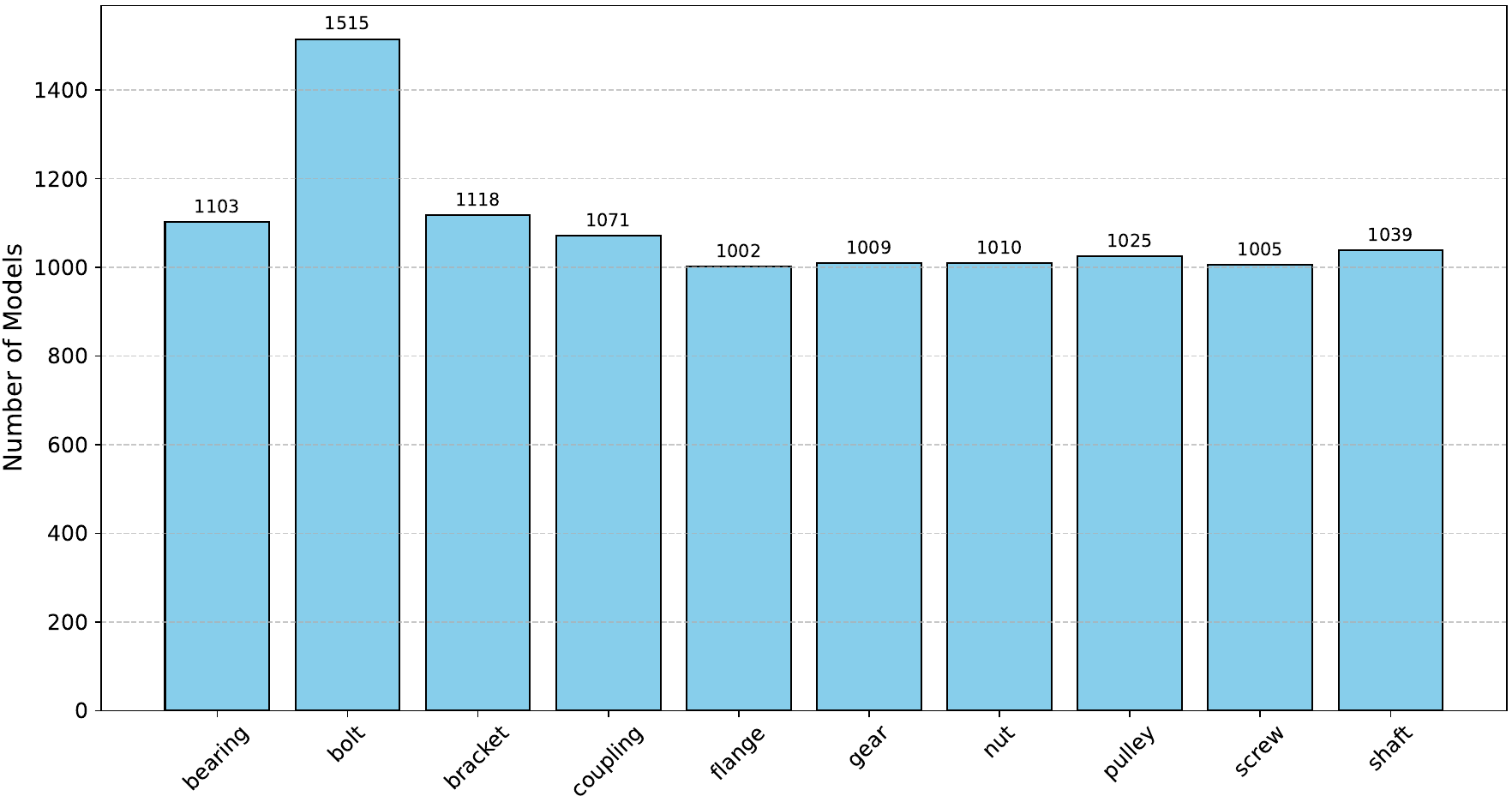}
    \caption{Per-class distribution of models in TMCAD.}
    \label{fig:mech_class_dist}
\end{figure*}

TMCAD stands for truly mechanical CAD. We collected real-world mechanical CAD models from both experienced designers and online sources, constructing a dataset of over 10,000 B-rep models stored in the STEP file format. The data is split into 10 classes: \textit{Bearing}, \textit{Bolt}, \textit{Bracket}, \textit{Coupling}, \textit{Flange}, \textit{Gear}, \textit{Nut},   \textit{Pulley}, \textit{Screw}, and \textit{Shaft}. Fig.~\ref{fig:TMCAD} shows some examples from TMCAD datasets. Fig.~\ref{fig:mech_class_dist} shows the distribution of models in different classes; there is an average
of 1,090 models per class in the dataset. The dataset is available at: https://github.com/Qiang-Zou/BRT.

\end{document}